\documentclass[aip,reprint,twocolumn,showpacs,preprintnumbers,amsmath,amssymb]{revtex4}

\usepackage{graphicx}
\usepackage{dcolumn}
\usepackage{bm}
\usepackage{psfrag}
\usepackage{color}
\usepackage{paralist}

\newcommand{\vex}[1]{\mathbf{#1}}
\newcommand{\vu}{\vex{u}}

\newcommand{\ten}[1]{\mathbf{#1}}
\newcommand{\tT}{\ten{T}}

\newcommand{\vnabla}{\boldsymbol{\nabla}}
\newcommand{\pdert}[1]{\partial_t #1}

\newcommand{\eqnref}[1]{Eq.~(\ref{#1})}

\def\square{${\vcenter{\hrule height .8pt
        \hbox{\vrule width .8pt height 5pt \kern 5pt
        \vrule width .8pt}
        \hrule height .8pt}}$}

\newcount\ndots
\def\drawline#1#2{\raise 2.5pt\vbox{\hrule width #1pt height #2pt}}

\def\trian{\raise 1.25pt\hbox{$\scriptscriptstyle\triangle$}\nobreak\ }
\def\circle{$\circ$\nobreak\ }

\def\square{${\vcenter{\hrule height .4pt
        \hbox{\vrule width .4pt height 3pt \kern 3pt
        \vrule width .4pt}
        \hrule height .4pt}}$\nobreak\ }
\def\plus{\raise 1.25pt \hbox{$\scriptscriptstyle +$}\nobreak\ }
\def\x{\raise 1.25pt \hbox{$\scriptscriptstyle \times$}\nobreak\ }

\def\solidtrian{\raise 1.25pt
   \hbox to 3bp{
\hfill}\nobreak\ }
\def\solidsquare{\vrule height .9ex width .8ex depth -.1ex\nobreak\ }

\def\solidcclose{\drawline{10}{.5}\nobreak\raise
  0.5pt\hbox{$\bullet$}\drawline{10}{.5}\nobreak\ }

\def\solidsclose{\drawline{10}{.5}\nobreak\raise
  0.5pt\hbox{\solidsquare}\drawline{10}{.5}\nobreak\ }

\def\solidtclose{\drawline{10}{.5}\nobreak\raise
  0.5pt\hbox{\solidtrian}\drawline{10}{.5}\nobreak\ }

\def\solidcopen{\drawline{10}{.5}\nobreak\raise
  0.5pt\hbox{\circle}\drawline{10}{.5}\nobreak\ }

\def\solidsopen{\drawline{10}{.5}\nobreak\raise
  0.5pt\hbox{\square}\drawline{10}{.5}\nobreak\ }

\def\solidtopen{\drawline{10}{.5}\nobreak\raise
  0.5pt\hbox{\trian}\drawline{10}{.5}\nobreak\ }

\def\solidx{\drawline{10}{.5}\nobreak\raise
  0.5pt\hbox{\x}\drawline{10}{.5}\nobreak\ }

\begin{document}

\title{Dynamics of Elasto-Inertial Turbulence}




\author{Y. Dubief}
\affiliation{School of Engineering, University of Vermont, Burlington VT}
\author{V. E. Terrapon}
\affiliation{Aerospace and Mechanical Engineering Department, University of Li\`{e}ge, Belgium}%
\author{J. Soria}
\affiliation{Department of Mechanical and Aerospace Engineering, Monash University, Australia\\ 
and Department of Aeronautical Engineering, King Abdulaziz University, Jeddah, Kingdom of Saudi Arabia}
  
\date{\today}

\maketitle
\maketitle



This movie illustrates the recent numerical and experimental discovery of a new state of turbulence in dilute polymer solutions, Elasto-Inertial Turbulence (hereafter referred to as EIT) reported in \cite{samanta2012transition, dubief2010polymer}. EIT is characterized by a chaotic flow state in which both inertial and elastic effects, whose relative contributions vary with the Reynolds number, control the flow dynamics. EIT explains the phenomenon of early turbulence \cite{hoyt1977laminar}, which describes the onset of turbulence in the presence of diluted polymer additives at Reynolds numbers significantly smaller than in the absence of polymers. EIT also offers a new perspective on the asymptotic drag reduction state in large Reynolds number polymer flows, maximum drag reduction or MDR \cite{virk1970uaa} in the limit of very large elasticity of the polymer flow. Indeed polymer additives are known for producing upward of 80\% of drag reduction in turbulent wall-bounded flows through a strong alteration and reduction of the turbulent activity \cite{white2008map}.

In the present movie, a periodic channel flow with carefully chosen initial conditions demonstrate the existence of EIT over a large range of Reynolds numbers. As shown in \cite{samanta2012transition}, the simulated evolution of the friction factor as a function of the Reynolds numbers matches a comparable experiment in a pipe flow.

The drag reducing mechanism is caused by an increase of the (extensional) viscosity in extensional upwash and downwash flows generated by quasi-streamwise vortices \cite{dubief2004cdr,terrapon2004sps}, thereby creating a negative torque on these near-wall vortices \cite{kim2007effects}. \cite{dubief2004cdr} demonstrated that polymers re-inject part of the energy accumulated  in high speed streaks, regions of locally high speed flow in the near-wall region elongated in the direction of the flow. In inertia-less flows with curved streamlines, \cite{groisman2000elastic} demonstrated the existence of strong non-linear mixing supported by elastic turbulence, a state of saturated dynamical interactions between stretched polymer molecules and the base flow that causes the stretching. EIT exists by either creating its own extensional flow patterns, as the movie demonstrates in subcritical channel flows and high elasticity flows.

Channel flow simulations are performed in a cartesian domain, where $x$, $y$
and $z$ are the streamwise, wall-normal and spanwise directions,
respectively. For a polymer solution, the flow transport
equations are the  conservation of mass, $\vnabla\cdot\vex{u}=0$, where $\vu$ is the velocity vector,
and transport of momentum:
\begin{equation}
\pdert{\vex{u}}+(\vex{u}\cdot\vnabla)\vex{u}=-\vnabla p+\frac{\beta}{Re}\nabla^2\vex{u}+\frac{1-\beta}{Re}\vnabla\cdot\tT\,. \label{eq:mom}
\end{equation}
The Reynolds number is based on the bulk velocity $U_b$ and the full channel height $H=2h$, $Re=U_bH/\nu$.
The parameter $\beta$ is the ratio of solvent viscosity to the zero-shear viscosity of the polymer solution and affects both the viscous stress and polymer stress terms in \eqnref{eq:mom}. The polymer stress tensor $\ten{T}$ is computed using the FENE-P (Finite Elastic Non-linear Extensibility-Peterlin) model \cite{bird1987dynamics}:
\begin{equation}
\ten{T}=\frac{1}{Wi}\left(\frac{\ten{C}}{1-\text{tr}(\ten{C})/L^2}-\ten{I}\right)\label{eq:taup}\;,
\end{equation}
where the tensor $\ten{C}$ is the local conformation tensor of the polymer solution and $\ten{I}$ is the unit tensor. The properties of the polymer solution are $\beta$, the maximum polymer extension $L$, and the Weissenberg number $Wi$ based on the solution relaxation time $\lambda$ and the flow time scale relevant to the dynamics of interest. Here $Wi$ is based on the wall shear-rate $\dot{\gamma}$ of the initial laminar flow at each $Re$, hence $Wi=\lambda\dot{\gamma}$.  The FENE-P model assumes that polymers may be represented by a pair of beads connected by a nonlinear spring defined by the end-to-end vector $\vex{q}$. The conformation tensor is the phase-average of the tensorial product of the end-to-end vector $\vex{q}$ with itself, $\ten{C}=\langle\vex{q}\otimes\vex{q}\rangle$, whose transport equation is
\begin{equation}
\pdert{\ten{C}}+(\vex{u}\cdot\vnabla)\ten{C}=\ten{C}(\vnabla\,\vex{u})+(\vnabla\,\vex{u})^\text{T}\ten{C}-\tT\;. \label{eq:C}
\end{equation}
On the right hand side of \eqnref{eq:C}, the first two terms are
responsible for the stretching of polymers by hydrodynamic forces,
whereas the third term models the internal energy that tends to
bring stretched polymers to their least energetic state (coiled).

Eqs.~(\ref{eq:mom}-\ref{eq:C}) are solved using finite differences
on a staggered grid and a semi-implicit time advancement scheme described elsewhere \cite{dubief2005nai}. A series of simulations was carried out for Reynolds numbers ranging from 1000 to 6000. A thorough resolution study  led to choose a domain size of $10H\times H\times 5H$ with $256\times151\times256$ computational nodes. All results discussed here have been verified on domains with a factor 2 in horizontal dimensions and resolution in each directions. The CFL number was set to 0.15  to guarantee the boundedness of $\ten{C}$.   


 The protocol for our simulations was designed to mimic the perturbed experimental setup of \cite{samanta2012transition} within the limitation inherent to the DNS boundary conditions. For any flow, Newtonian or polymeric, the initial flow and polymer fields are first equilibrated to the laminar state corresponding to the desired $Re$.  A perturbation is then introduced over a short duration, in the form of blowing and suction velocity on both walls, over which white noise of prescribed intensity is introduced. The velocity pattern is periodic in $x$ and $z$:
 \begin{equation}
 v_w(x,z,t)={\cal H}(t)\left[A\sin\left(\frac{8\pi}{L_x} x\right)\sin\left(\frac{8\pi}{L_z}z\right)+\varepsilon(t)\right],\label{eq:vw}
 \end{equation}
 where A is the amplitude, $L_x$ and $L_z$ are the horizontal domain dimensions, and $\varepsilon(t)$ is the random noise. The total duration of the perturbation is $0.5h/U_b$, of which the first and last 10\% correspond to a gradual increase / decrease through a smooth step function ${\cal H}(t)$. Choosing $A=0.09U_b$ and the RMS of $\varepsilon$ at $0.005U_b$ causes the Newtonian flow to transition at $Re=6000$.

All polymeric simulations are performed with $L=200$ and $\beta=0.9$. Two Weissenberg numbers are considered, $Wi=100$ and 700. The former is consistent with previous simulations of MDR \cite{dubief2004cdr,li2006influence,white2012re}.  \cite{procaccia2008colloquium}'s theory, based on infinite $Re$ and $Wi$, motivates the second, as an exploration of the effects of very large elasticity, or $Wi$, on the flow.

This work was performed during the 2012 Center for Turbulence Research Summer Program and will be published in \cite{dubief2012dynamics}.
\begin{acknowledgments}
The Vermont Advanced Computing Center is gratefully acknowledged for providing the computing resources necessary for our simulations. YD acknowledges the support of grant No. P01HL46703 (project 1) from the National Institutes of Health. VET acknowledges the financial support of a Marie Curie FP7 Career Integration Grant within the 7th European Community Framework Programme (Grant Agreement n$^\circ$ PCIG10-GA-2011-304073). JS acknowledges the support of the Australian Research Council.
\end{acknowledgments}

\bibliography{ctr_summer_dubief}

\end{document}